\documentclass[a4paper,oneside,12pt]{article}
\pdfoutput=1 

\usepackage[textwidth=17cm,textheight=24cm]{geometry}
\usepackage{hyperref}
\usepackage{authblk}
\usepackage[pdftex]{graphicx}
\usepackage[font=small]{caption}   

\usepackage{placeins}

\providecommand{\keywords}[1]{\textbf{\textit{Keywords---}} #1}

\begin{document}

This is an author-created, un-copyedited version of an article accepted for publication in JINST. 
IOP Publishing Ltd is not responsible for any errors or omissions 
in this version of the manuscript or any version derived from it. 
The Version of Record is available online at \url{https://doi.org/10.1088//1748-0221/15/09/c09059}.

\title{\boldmath The PANDA DIRCs}


\author[a]{C.~Schwarz,\footnote{Corresponding author.}}
\author[a,b]{A.~Ali,}
\author[a]{A.~Belias,}
\author[a]{R.~Dzhygadlo,}
\author[a]{A.~Gerhardt,}
\author[a,b]{M.~Krebs,}
\author[a]{D.~Lehmann,}
\author[a,b]{K.~Peters,}
\author[a]{G.~Schepers,}
\author[a]{J.~Schwiening,}
\author[a]{M.~Traxler,}
\author[c]{L.~Schmitt,}
\author[d]{M.~B\"{o}hm,}
\author[d]{A.~Lehmann,}
\author[d]{M.~Pfaffinger,}
\author[d]{S.~Stelter,}
\author[e]{M.~D\"{u}ren,}
\author[e]{E.~Etzelm\"{u}ller,}
\author[e]{K.~F\"{o}hl,}
\author[e]{A.~Hayrapetyan,}
\author[e,a]{I.~K\"{o}seoglu,}
\author[e]{K.~Kreutzfeld,}
\author[e]{M.~Schmidt,}
\author[e]{T.~Wasem,}
\author[f]{C.~Sfienti,}
\author[g]{A.~Barnyakov,}
\author[g]{K.~Beloborodov,}
\author[g]{V.~Blinov,}
\author[g]{S.~Kononov,}
\author[g]{E.~Kravchenko,}
\author[g]{and I.~Kuyanov}
\affil[a]{GSI Helmholtzzentrum f\"ur Schwerionenforschung GmbH , Darmstadt, Germany}
\affil[b]{Goethe University, Frankfurt a.M., Germany}
\affil[c]{FAIR, Facility for Antiproton and Ion Research in Europe, Darmstadt, Germany}
\affil[d]{Friedrich Alexander-University of Erlangen-Nuremberg, Erlangen, Germany}
\affil[e]{II. Physikalisches Institut, Justus Liebig-University of Giessen, Giessen, Germany}
\affil[f]{Institut f\"{u}r Kernphysik, Johannes Gutenberg-University of Mainz, Mainz, Germany}
\affil[g]{Budker Institute of Nuclear Physics of Siberian Branch Russian Academy of Sciences, Novosibirsk, Russia}

\maketitle
\flushbottom

\abstract{
The PANDA experiment at the FAIR facility adresses open questions 
in hadron physics with antiproton beams in the momentum range of
1.5-15 GeV/$c$. The antiprotons are stored and cooled in a High Energy
Storage RING (HESR) with a momentum spread down to
\mbox{$\Delta$p/p = 4$\cdot10^{-5}$}. A  high luminosity
of up to $2 \cdot 10^{32} $cm$^{-2}s^{-1}$ can be achieved.
An excellent hadronic particle identification (PID) will be provided
by two Cherenkov detectors using the priciple of
Detection of Internally Reflected Cherenkov light (DIRC). In the forward
direction from polar angles of   $5^\circ$ to $22^\circ$, the Endcap Disc DIRC (EDD)
separates pions from kaons up to momenta of 4 GeV/$c$. Between
 $22^\circ$ and $140^\circ$ the Barrel DIRC
cleanly separates pions from kaons for momenta up to \mbox{3.5 GeV/$c$}.
This article describes the design of the Barrel DIRC and of the Endcap Disc DIRC
and the validation of their designs in particle beams at the CERN PS.
}
\vspace{5mm}



\section{Introduction}
\label{sec:intro}

The Facility for Antiproton and Ion Research (FAIR) close to GSI Darmstadt in Germany is currently
under construction. The tunnel for the synchrotron is being dug out and the concrete sections
are being poured. 
Four large experiment collaborations are the scientific pillars of the project. Three of them, CBM,
NUSTAR, and APPA will pursue the traditional physics of GSI Darmstadt with heavy ion beams. The fourth,
PANDA (antiProton ANihilations at DArmstadt), will benefit from a new available antiproton production chain.
Antiproton beams with unprecedented intensity and quality are stored and accelerated in the
High Energy Storage Ring (HESR) in the  momentum range between 1.5 to 15 GeV/$c$ and 
annihilate with a fixed target.
The experiments with charmed quarks will shed light on strong QCD in a beam momentum range where  
perturbation theory is still valid, but the influence of strong QCD becomes visible.
The accumulation of up to $10^{11}$ antiprotons yields a  luminosity of up to $2 \cdot 10^{32} $cm$^{-2}s^{-1}$.
The momentum resolution of the beam is improved by stochastical
cooling to \mbox{$\Delta$p/p = 4$\cdot10^{-5}$} and enables high precision 
spectroscopy.
\begin{figure}[hb]
\captionsetup{width=0.8\textwidth}
\centering 
\includegraphics[width=.8\textwidth,trim=0 0 0 0,clip]{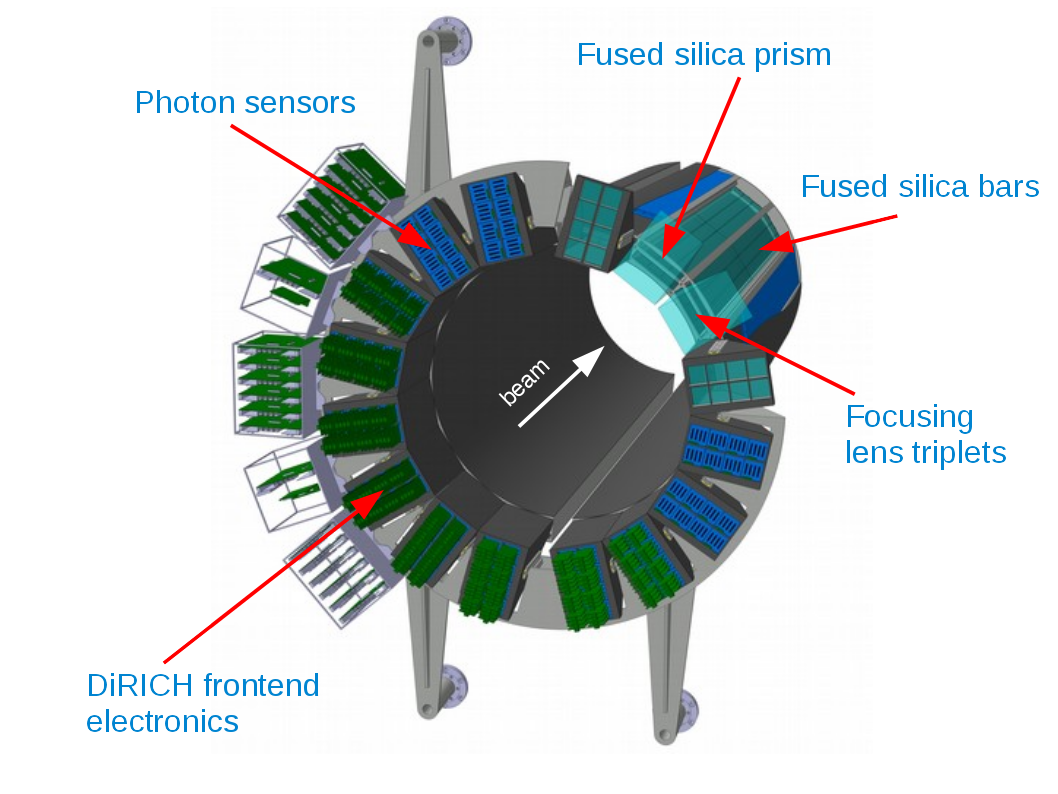}
\caption{\label{fig:1} Schematic of the PANDA Barrel DIRC.}
\end{figure}
Here, resonances produced in formation experiments are scanned by beams with different momenta and
their widths can be determined with an energy resolution in the order of  \mbox{$\Delta$E $\approx$ 50~keV}.
Above the threshold for open charmed mesons the identification of kaons
with Ring Imaging CHerenkov counters (RICH) becomes an important task for the PANDA experiment.
The detector consists of two parts, the target spectrometer around the fixed
target and the forward spectrometer for the measurement of the forward boosted reaction products.
In the forward spectrometer a focusing aerogel RICH \cite{Kononov19} provides charged PID.
The target spectrometer contains two RICH
counters.
\begin{figure}[bht]
\captionsetup{width=0.8\textwidth}
\centering
\includegraphics[width=.4\textwidth,trim=0 0 0 0,clip]{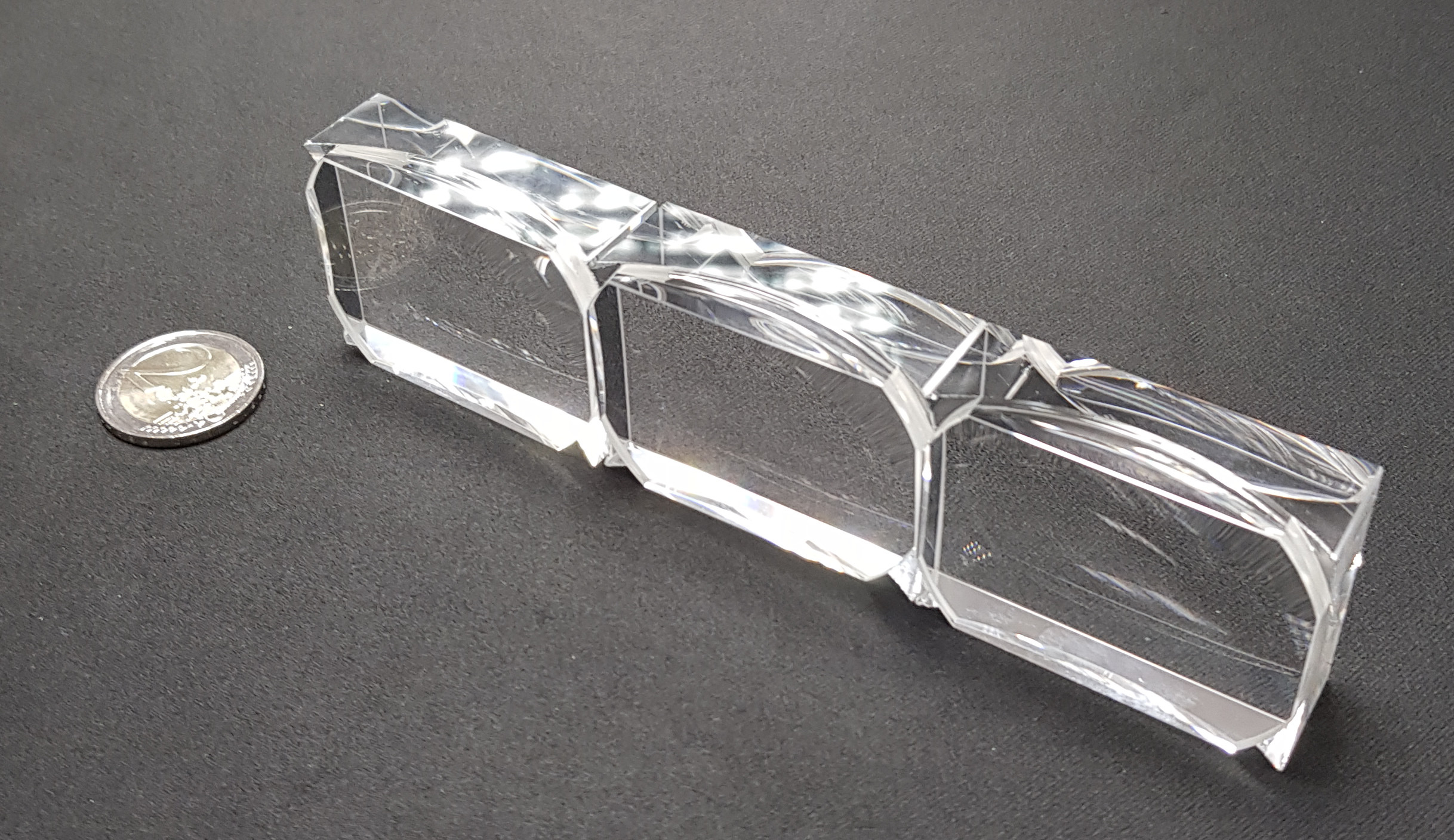}
\caption{\label{fig:2} Three focusing lens triplets built by Befort Wetzlar Optics \cite{Befort},
each being a lanthanum crown glass lens between two fused silica pieces with
flat outer surfaces. They focus the light from the radiator bar on the back side of the prism.}
\end{figure}
They have to fit into the space inside of a surrounding lead tungsten calorimeter.
The required space of a traditional RICH leads to an increase in size and costs of the calorimeter. 
Therefore, the DIRC principle was chosen. Here, the radiator
acts also as a ligthguide and preserves the information of the Cherenkov angle of the photons of the charged particle
over many internal reflections. From polar angles between $22^\circ$ and $140^\circ$ the Barrel DIRC will
distinguish kaons from pions up to momenta  of  \mbox{3.5 GeV/$c$} with a separation power of 3 standard deviations (s.d.)
 \cite{BarrelTDR,Schwiening18,Schwarz20}.  
 In the forward direction from polar angles of $5^\circ$ to $22^\circ$ the Endcap Disc DIRC (EDD)
 \cite{Etzelmueller19,Koeseoglu20}
provides pion-kaon separation of 3 s.d. up to momenta of  \mbox{4 GeV/$c$}. The requirements for the kaon momenta
come from GEANT4 \cite{geant4}  simulations of the phase space for several reaction channels.
The following sections describe the design of the DIRCs and the results of experiments in test beams.

\section{The Design}
\label{sec:design}

The successfully operated BaBar DIRC and the development work for the SuperB FDIRC inspired the design
of the PANDA Barrel DIRC (Figure \ref{fig:1}).
An improvement is the  smaller readout volume, made from synthetic fused silica,
which is easier to integrate and is less affected by unwanted radiation coming from the accelerator.
It is located within the yoke of the solenoid magnet of the target spectrometer and is divided in 16 prisms,
300 mm long with an opening angle of  $33^\circ$.
An array of $2\times4$ pixelated Microchannel Plate Photomultiplier Tubes (MCP-PMTs) attached to the upstream end of the prism
using a RTV-silicone cookie
allows the operation of the readout within a magnetic field of 1 Tesla.
The tubes are expected to survive 10 years of the operation of the PANDA experiment \cite{lehmann:mcp}.
Each MCP-PMT has an $8\times8$ anode grid with a pixel pitch of $6.5$~mm.
Due to the fast detector response and the
frontend electronics using the DiRICH system \cite{traxler:dirich} a timing precision of 100 ps is
achievable \cite{BarrelTDR,Schwiening18,Schwarz20}.
Each prism matches a bar box with three radiator bars made from synthetic fused silica. The bars are 
17 mm thick, 53 mm wide, and 2400 mm long. 
One radiator bar consists of two 1200 mm long pieces glued together
end-to-end. Photons hitting the downstream end of the radiator are reflected back by a mirror towards the readout
volume. For a precise angular measurement of the photons from the large cross section of the radiator bars
focusing optics is needed. Space limitations and the positioning of the photodetectors in the magnetic field
favor the use of a lens system. 
A picture of three lens triplets are shown in Figure \ref{fig:2}.
The lens is glued to the upstream end of the bar and a RTV-silicone cookie couples the lens system
to the prism.

\begin{figure}[tbh]
\captionsetup{width=0.8\textwidth}
\centering
\includegraphics[width=.85\textwidth,trim=0 0 0 0,clip]{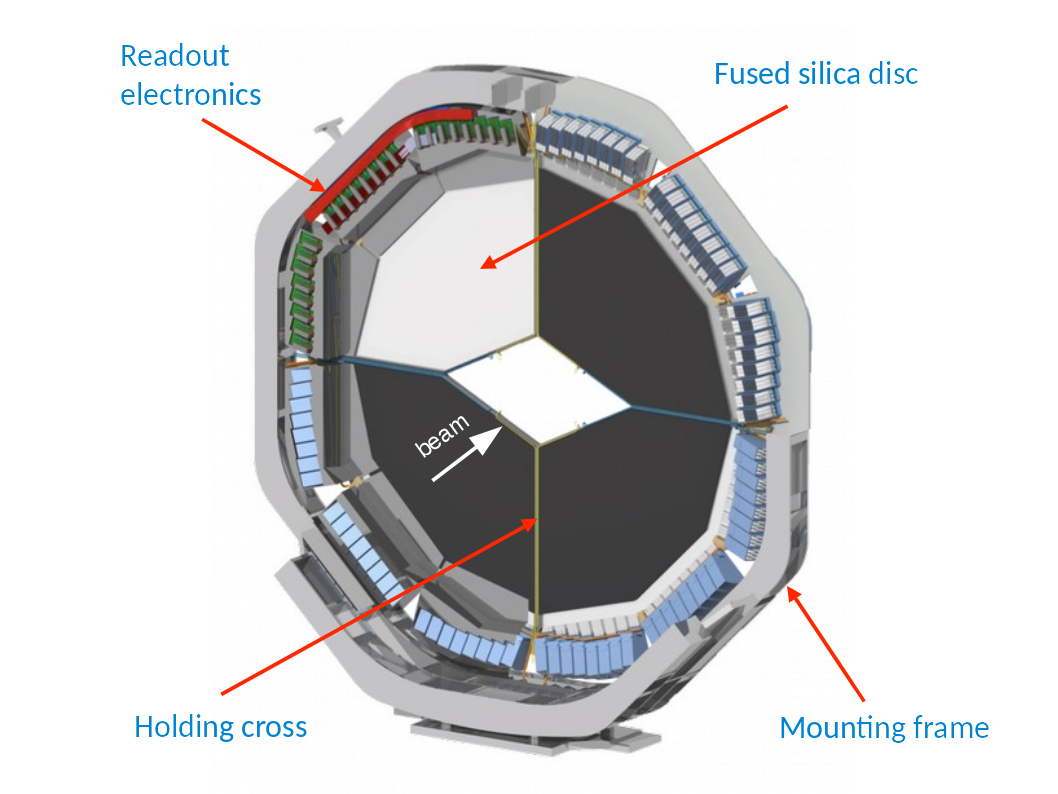}
\caption{\label{fig:3} Schematic of the Endcap Disc DIRC.}
\end{figure}

\begin{figure}[bth]
\centering 
\includegraphics[width=.95\textwidth,trim=0 0 0 0,clip]{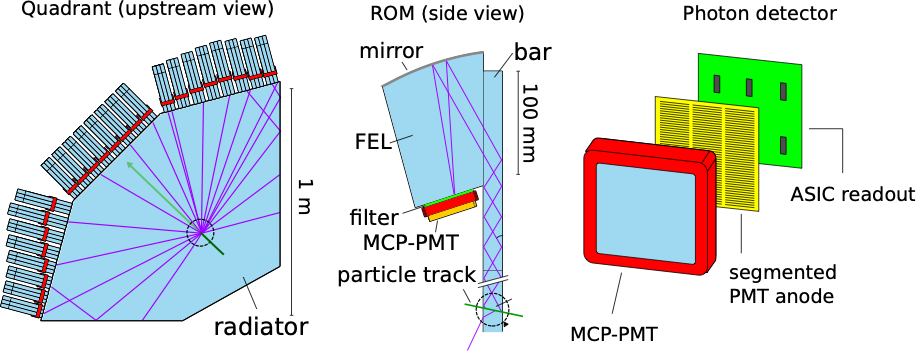}
\caption{\label{fig:4} Schematic of one EDD quadrant and its readout.}
\end{figure}

While the design of the Barrel Dirc utilizes the experience from the BaBar DIRC and the SuperB FDIRC,
the Endcap Disc DIRC is a novel approach.
The EDD uses four 20~mm-thick fused silica radiator plates forming a disc (Figure \ref{fig:3}) in front of a
lead tungstate forward endcap electromagnetic calorimeter.
The reflections of the photons at the polished faces of the plate
conserve the internally reflected Cherenkov angle during their propagation towards the outer rim. 
The sides of the outer rim are equipped 
with focusing elements and MCP-PMTs as photon sensors.
The four quadrants are optically isolated and the photons reflected on the interior sides. 
The quadrants are stabilized by a holding cross mounted to a ring shaped mounting frame.
The EDD and the forward endcap electromagnetic calorimeter are planned to be fixed on opposite sides of the same
mounting frame. 
Each quadrant has 24 readout
modules (ROM) with attached MCP-PMTs. Each ROM consists of 3 focusing elements (FELs) as expansion volumes
with a cylindrical mirror at the backside \mbox{(Figure \ref{fig:4})}. 
The EDD in total consists of 96 ROMs, 96 MCP-PMTs, and 288 FELs. 
The position of the hit FEL and the position of the charged particle on the radiator from the tracking system
determine the azimuthal
direction of the photon. The polar angle of the photon is measured by the highly segmented anode of the MCP-PMT with
3 $\times$ 100 pixels.
An ASIC read out board is connected to the MCP-PMT anode.
This readout board of the EDD is being developed by PETsys \cite{PETsys}.

\section{Experiments with Test Beams}

Since 2008 various PANDA DIRC prototypes have been evaluated with particle beams at CERN, DESY, and GSI.
Here, we focus on the test campaign
at the CERN-PS in 2018, when the Barrel DIRC and the EDD prototypes were installed in the T9 beamline.
The mixed hadron beam contained primarily pions and protons, which were cleanly tagged by a time-of-flight system.
For the Barrel DIRC most runs were taken at a momentum of 7 GeV/$c$. At this momentum the difference in
the Cherenkov angle between protons and pions is close to that between kaons and pions at a momentum of 3.5 GeV/$c$,
which is the designed upper momentum limit of the Barrel DIRC.
In this campaign the possible reduction
of the number of MCP-PMTs for cost optimization, different optical couplings, and housing and cable
\begin{figure}[bh]
\captionsetup{width=0.8\textwidth}
\centering 
\includegraphics[width=.95\textwidth,trim=0 0 0 0,clip]{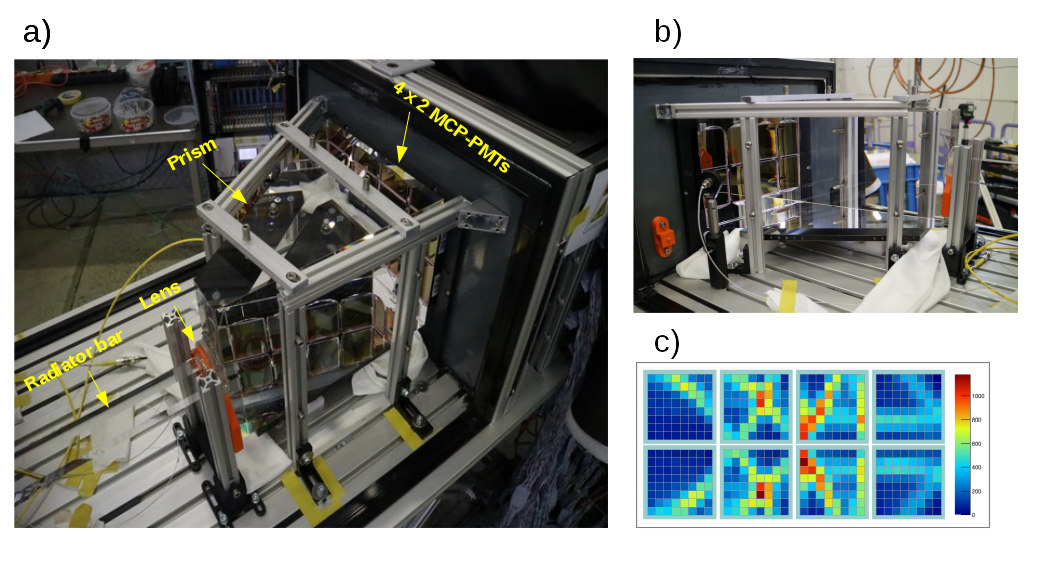}
\caption{\label{fig:5} The Barrel DIRC setup from top (a) and the side (b). The colored histogram (c) shows
the accumulated hit pattern for a polar angle
of $20^\circ$.}
\end{figure}
routing of the frontend electronics were studied. The setup with a 35 mm wide radiator bar
is shown in Figure \ref{fig:5} a) and b).
A radiator bar, made from synthetic fused silica, was mounted on a turntable. By rotation of the turntable the
polar and azimuthal angle between the radiator and the beam could be adjusted. Cherenkov photons going to the
left end of the bar in Figure \ref{fig:5} a) were reflected by a flat mirror towards the right side of the bar.
Here, a three-layer spherical lens couples the bar to a 300 mm deep prism. The middle part of the  lens is
made from N-LaK33B glass \cite{schott} and is sandwiched between two synthetic fused silica pieces. 
It has a defocusing and a focusing surface to yield a sufficiently flat focal plane on the backside of the prism, equipped with an
 \mbox{2 $\times$ 4} matrix of MCP-PMTs.
A side view of the prism with the MCP-PMT array is shown in Figure  \ref{fig:5} b). 
In the beam tests the MCP-PMTs XP85012/A1-Q from PHOTONIS \cite{photonis} were used.
The colored histogram in Figure \ref{fig:5} c) shows
the accumulated hit pattern of Cherenkov photons for
beam particles with a momentum of 7 GeV/$c$ hitting the bar at a polar angle of $20^\circ$.
The electronics for the readout was based on the Trigger Readout Board version 3 (TRB3) of the
HADES collaboration \cite{trb3-jinst} and measures the time of arrival and the time over threshold
of logical signals coming from PADIWA discriminator boards \cite{cardinali:padiwa} plugged onto the MCP-PMTs. 
The timing offsets and the timing precision of the full readout chain were determined 
using a 405 nm Picosecond Injection Laser (PiLas) \cite{pilas}.
The measured timing precison showed on average values of 190 ps after walk correction.
\begin{figure}[tb]
\captionsetup{width=0.8\textwidth}
\centering 
\includegraphics[width=.75\textwidth,trim=0 0 0 0,clip]{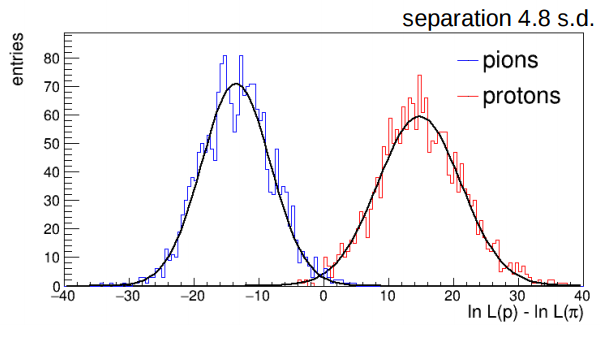}
\caption{\label{fig:6} Difference of the maximum likelihood for the arrival time of photons from protons/pions at 7 GeV/$c$
  (equivalent to kaons/pions at 3.5 GeV/$c$) at a polar angle of $20^\circ$.}
\end{figure}
\begin{figure}[bth]
\captionsetup{width=0.8\textwidth}
\centering 
\includegraphics[width=.99\textwidth,trim=0 0 0 0,clip]{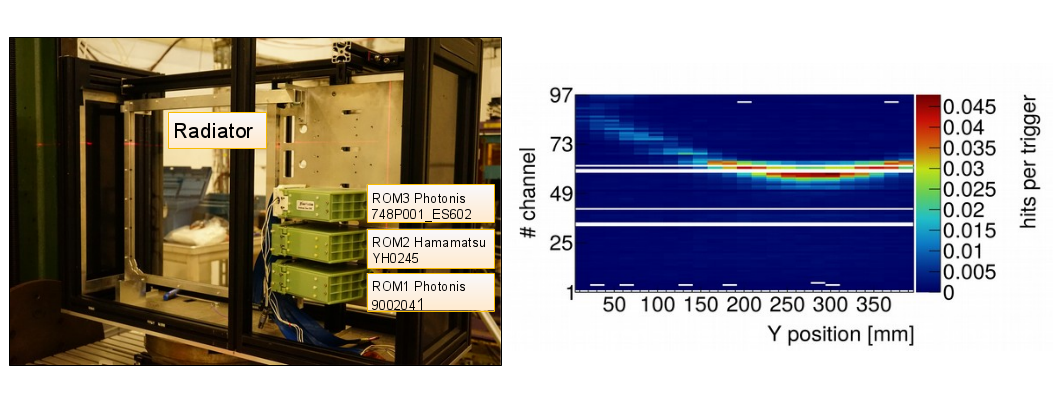}
\caption{\label{fig:7} The setup of the Endcap Disc DIRC (left) is a radiator plate with with 3 ROMs attached.
  Each ROM contains three focussing lightguides read out by one pixelated MCP-PMT. The prototype could be moved horizontal
  and vertical for position scans and rotated around the vertical axis for angular scans. 
The colored histogram (right) shows the accumulated hit pattern for many events.}
\end{figure}
There are two reconstruction algorithms to determine the performance of the detector \cite{Ali20,Dzhygadlo19}.
The "geometrical reconstruction" uses the position of the detected photons to make a
track-by-track maximum likelihood fit.   The
"time imaging" employs both, the position and time of arrival of detected photons to perform directly the maximum likelihood fit. 
The difference of the maximum likelihood  using the time imaging of photons for protons and pions
is shown in Figure \ref{fig:6}.
The separation power for protons and pions of 4.8 s.d. at a momentum of 7 GeV/$c$
and a polar angle of  $20^\circ$ clearly exceeds the PANDA PID requirements.

The Endcap Disc DIRC prototype used a synthetic fused silica plate ($500 \times 500 \times 20$~mm$^3$)
placed on a vertical and horizontal adjustable table. In addition, the prototype could be rotated arround the vertical axis.
During the beam test several vertical and horizontal position scans and angular scans were performed. 
One edge  of the radiator plate was 
equipped with 3 ROMs (9 FELs) and different photon detectors from PHOTONIS\cite{photonis} and HAMAMATSU \cite{hamamatsu}
(Figure \ref{fig:7}, left).
The opposite edge of the radiator was connected to a PiLas laser to calibrate time differences between the electronic channels.
In this campaign 
the version of the TOFPET ASIC, which was optimized for positive signals from silicon photomultiplier,
showed problems with the negative polarity from the MCP-PMTs, resulting in a loss of hit detection efficiency.
The problem is understood and a new ASIC version is in production.
The number of detected Cherenkov photons per trigger as a function of beam position and 
channel number at 7 GeV/$c$ for pions is shown on the right side in Figure  \ref{fig:7}. The channel number and the
beam position are a measure for the polar and azimuthal angle of the Cherenkov photons, respectively. 
\begin{figure}[bh]
\centering 
\includegraphics[width=.75\textwidth,trim=0 0 0 0,clip]{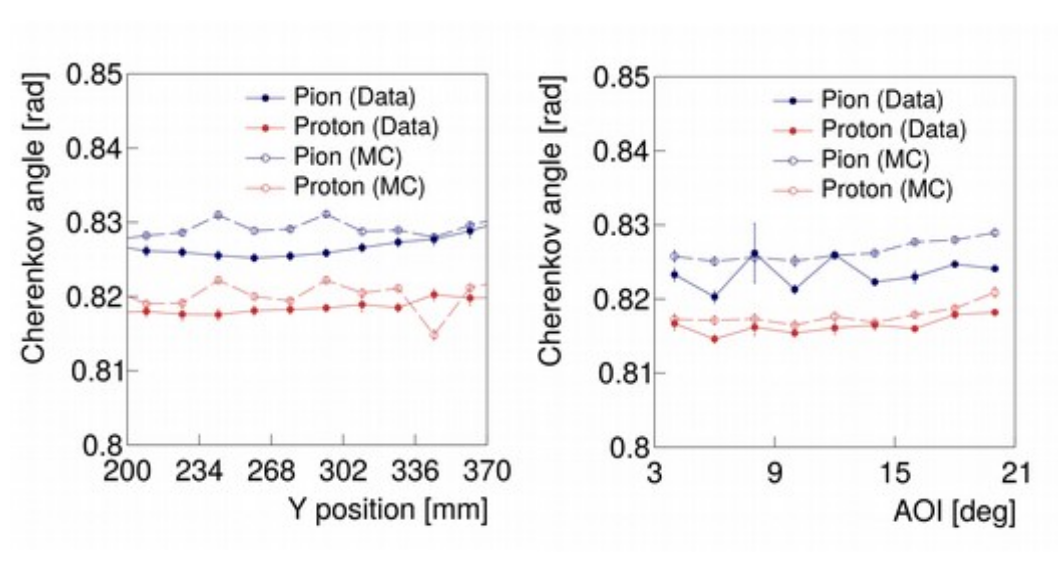}
\caption{\label{fig:8} Reconstructed Cherenkov angle for different vertical positions (left) and 
for different angles of the incident beam (right) are compared to predictions from GEANT4 simulations.}
\end{figure}The Endcap Disc DIRC prototype used a synthetic fused silica plate ($500 \times 500 \times 20$~mm$^3$)
Each measured Cherenkov hit was reconstructed by using a geometrical reconstruction algorithm \cite{Schmidt17}. 
The results of the position (Figure \ref{fig:8}, left) and angular scans (Figure \ref{fig:8}, right) ,
first published in  \cite{Koeseoglu20}, agree with expectations from GEANT4 simulations.

\section{Outlook}
Following the successful performance validation with particle beams
and the completion of the technical design reports the DIRCs have entered the construction phase.
The Barrel DIRC components with the longest production time, 
the radiators, have been ordered from Nikon \cite{nikon}. The tendering of the MCP-PMTs 
is at an advanced stage. For the EDD the component fabrication will start.
However, there are still important R\&D topics: 
The latest generation of the Barrel DIRC readout electronics, the DiRICH system \cite{traxler:dirich},
originally designed for multianode-PMTs, needs
the adaption of the discriminator input stage to the fast signals of MCP-PMTs.  
The coupling between the bar boxes and prisms and between the MCP-PMTs and the prism 
will be done with RTV-silicone cookies and is subject of
ongoing tests. The material of the bar boxes can be aluminum or  carbon fiber reinforced polymer (CFRP).
Long-term outgassing tests of CFRP and its effect on the surfaces of the radiator will show
if this material is better to minimize the material budget.
The next generation of the TOFPET ASIC for the EDD compatible with both polarities
will be tested in the Giessen Cosmic Station (GCS) \cite{Bodenschatz20}.

The experimental hall for the PANDA detector will be ready to move in and to install first basic elements,
like the solenoid, in 2022. The Barrel DIRC will then be installed in 2023/2024 and subsequently commissioned.
At the same time one quadrant of the the EDD will be installed as a prototype followed by a complete installation in 2026/27.


\section*{Acknowledgments}

This work was supported by 
HGS-HIRe, 
HIC 
for FAIR.
We thank the CERN staff for the opportunity to use 
the beam facilities and for their on-site support.


\end{document}